\documentclass{aastex631}

\begin{document}

\title{Species Syzygy: Which Animal Has Seen the Most Total Solar Eclipses?}

\shorttitle{\sc{Species Syzygy}}
\date{April 1st 2024}

\correspondingauthor{Mark Popinchalk}
\author[0000-0001-9482-7794]{Mark Popinchalk}
\affiliation{\amnh}
\email{popinchalkmark@gmail.com}

\newcommand{\amnh}{Department of Astrophysics, American Museum of Natural History, Central Park West at 79th Street, New York, NY 10034, USA}

\begin{abstract}
A total solar eclipse (TSE) is a shocking and sublime experience. In just a week hundreds of millions of Homo Sapiens will attempt to see the 2024 eclipse as it stretches across the North American continent. However, while Homo Sapiens may be uniquely positioned to understand and predict eclipses, they are not the only species capable of observing them. The precise alignment of the Moon, Earth and Sun all existed well before humans. In the same way we share this planet capable of hosting life, the fantastic astronomical experiences available on it are not exclusive either. We present a framework to calculate the number of Total Solar Eclipses experienced by a species at any point in Earth's history. This includes factoring in the evolution of the Sun-Moon-Earth system, the duration the species is extant, and average population. We normalize over the geographic range by calculating an Astronomical World Eclipse Surface cOverage MEtric (AWESOME) time. To illustrate this framework we look at the case study of the family \textit{Limulidae} (Horseshoe Crabs) and estimate the number of individuals that have seen an eclipse. We compare it to the number of current Homo Sapiens that view eclipses, and predict if it is possible for another species to take the ''top'' spot before the final total solar eclipse in $\sim$ 380 million years.
\end{abstract}

\section{Introduction}

A total solar eclipse (TSE) is when the alignment of the Earth, Moon and Sun is such that from a perspective on Earth, the disc of the Sun is completely obscured by the Moon. It is a geometric arrangement caused by the Moon and Sun having comparable angular sizes. It is a true marvel that this happens at all, let alone with regularity when you start to consider all the orbital variations involved. Apsidial precession of the Earth and the Moon, Nodal precession of the Moon, eccentricity variations of the Moon and Earth are all really challenging to keep track of, especially for an April Fools day paper.

In addition, the Earth moon system is evolving. The tidal effects of the Moon on the Earth create a torque on the Earth that both slows down the day and causes the Moon to slowly retreat. This change in orbit structure has been explored in other well regarded papers \citep{party}, but it's effect on Total Solar Eclipses has not been studied thoroughly.

There has been a lot of wildlife to witness those Total Solar Eclipses. While there is evidence of microbial life early in the Earth's history, it is near impossible to say much about them. However, the Cambrian Explosion $\sim$ 500 million years ago introduced a wide array of animals onto the surface of the Earth\cite{cambrian}.

Determining if an animal is aware of an eclipse can be challenging. Anecdotal stories of birds going to roost or performing night calls are common, but quantifying or defining ''observation'' from these stories is challenging. However, recently the study of animal reactions to total solar eclipses has been possible to study due to eclipse paths aligning with zoos in metropolitan areas. \cite{ZOO_Hartstone-Rose2020} tracked the response of 17 taxa of animals during thee 2017 TSE in Riverbanks Zoo and Garden in Columbia, South Carolina. 13 of the 17 acted differently from their baseline behavior, with 8 performing nighttime routines. The next most frequent behavior was anxiety, specifically noted in a few primates (both gorillas and baboons), matching the common reaction of humans prior to the last few centuries.

These are only the outward behaviors of the animals, and do not consider the inner workings of their minds. For example, \cite{ZOO_Hartstone-Rose2020} observed the Galapagos turtles turning to look skyward. Were they contemplating the eclipse? While there are contested reports of understanding animal communication \cite{doolittle}, we are unable to know for sure. Even if they don't understand what a Total Solar Eclipse is, they have borne witness. 

We therefore aspire in this paper to provide a framework for calculating how many members of a species have witnessed a Total Solar Eclipse. In Section~\ref{sec:animals} we present the biological considerations necessary to discussing animals witnessing TSE on astronomical timescales ($\sim$ 10 Myr). Section~\ref{sec:tse_time} we describe our method for calculating the number of TSE that will happen in a certain epoch in Earth's history. In Section~\ref{sec:calculate} we calculate the number of TSEs experienced by Horseshoe Crabs and by Humans as a species, and in Section~\ref{sec:disc} we discuss our findings and offer conclusions in Section~\ref{sec:con}.

\section{Lifespan of Species and Estimation of Animal Populations}\label{sec:animals}

We will start by taking a liberal approach to the definition of ''observing'' to include animals capable of distinguishing between light and dark, and carrying behaviors associated with each. Some might complain that this is too broad, but if you've ever been in the path of totality, you can't help but notice it. We hope that any critics of this choice will some day see the lack of light of day for themselves.

\subsection{Species Extinction Rate}

99.99$\%$ of all species have gone extinct\cite{extinction}. While this sounds like a bummer, it is a natural part of evolution. Most animals tend to exist on the order of 1-10 million years, and so that is the characteristic timescale we will consider. The exception being our case study Horseshoe Crabs (see Section~\ref{sec:horseshoe}).

\subsection{Geographic Range of Species}\label{sec:range} 

Animals obviously live in a variety of conditions and geographic regions on the Earth. Most are conducive to witnessing a TSE, as while they are varied they are all generally considered to be ''outside''. However, bodies of water represent something one can be inside of, and therefore we must account for the time spent in water for a given animal.

Attempting to calculate the number of eclipses for a specific location would be extremely challenging, especially since tectonic shift would move those locations or render them below the crust on the timescales we are concerned with. Thankfully, our method is independent of specific location, as you will see in Section~\ref{sec:awesome}.

\subsection{Animal Population and Anthropogenic Effects thereof}\label{sec:ani_pop}

To accurately gauge the number of TSEs members of a species have witnessed, the population of said species needs to be known. In particular, we need an estimate of the average population of a species over 10 Million year time spans. One issue for estimating this is that since humans have been able to count wild animals, they have been killing them.

This is not hyperbole. 

100,000 years ago, after modern humans existed \citep{humans}, there was 20 million tonnes of carbon in wild land mammals of which humans were negligible amount. By 2015 Humans consisted of 60 million tonnes and wild land mammals only 3 million \citep{owid-wild-mammals-birds-biomass}. Humans have without debate diminished populations and extinguished species, not only of mammals but across the animal kingdom. This makes present day population counts of our fellow species vast underestimates of any prior epoch.

\subsection{Case Study: Horseshoe Crabs}\label{sec:horseshoe}

We now apply the following considerations to a case study, Horseshoe Crabs. First and foremost, Horseshoe Crabs are capable of observing a Total Solar Eclipse. Horseshoe crabs have 10 total eyes, including two compound eyes and two sensitive to UV light, specifically used to track the Sun and reflected light from the Moon \citep{horseshoe_anatomy}.  

They spawn on the beach with the spring tide during the months of March and July \citep{horseshoe_mating_season}. The rest of the time they spend approximately 100 km offshore scavenging for food on the ocean floor \citep{horseshoe_range}. 

Most importantly, Horseshoe Crabs are considered living fossils. Since they emerged in the fossil record 480 million years ago, they have changed very little in function or form \citep{horseshoe_start}. In the last 200 million years they have essentially seen next to no evolution. There are currently 4 extant species, located on the east coasts of North America and Asia. While no single species can be traced back to 480 million years ago, to serve as a demonstration of our method we consider Horseshoe Crabs together, the family \textit{Limulidae}.

\begin{figure}
\begin{center}
\includegraphics[width=0.85\linewidth]{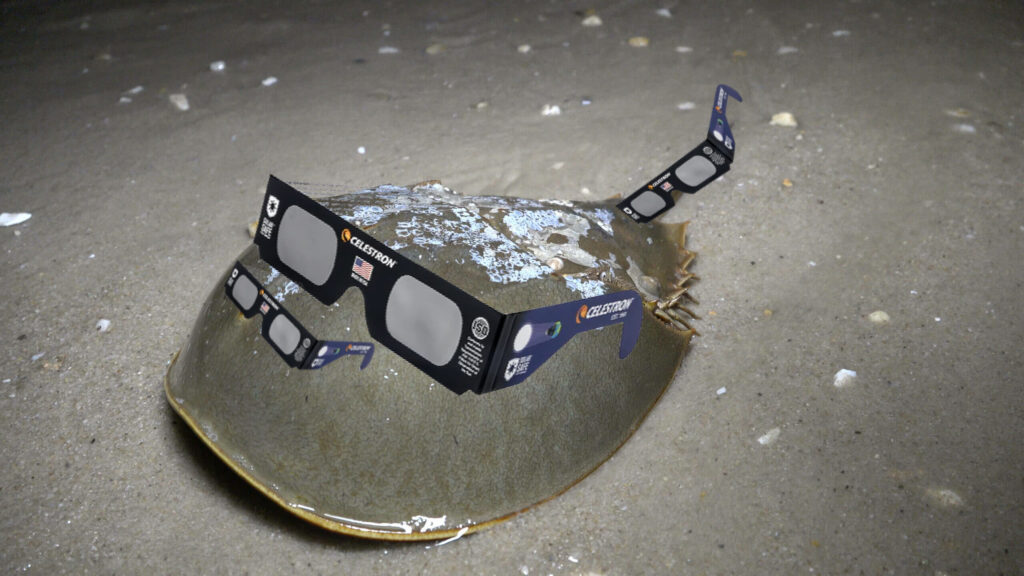}
\caption{An image of an Atlantic Horseshoe Crab, \textit{Limulus polyphemus}, wearing proper eye protection for viewing the Sun before a Total Solar Eclipse.}
\label{fig:horseshoe}
\end{center}
\end{figure}

In Figure~\ref{fig:horseshoe} we present a typical specimen of \textit{Limulus polyphemus}, the Atlantic Horseshoe Crab, wearing eclipse glasses on 8 of its 10 eyes.

\section{Frequency of Total Solar Eclipses over Earth's History}\label{sec:tse_time} 

As outlined in the introduction, a Total Solar Eclipse requires a precise alignment of the Earth Moon and Sun. However, while the orbital mechanics of the Earth Moon and Sun vary over time, we are going to sweep most of them under the rug. Precession rates, Saros cycles and eccentricity variations are all on the order of tens to hundreds of thousands of years. We will be working in 10 million year time steps, and so any effect due to the mentioned will be averaged out \citep[e.g.][]{inso_2004}.

We will state the average eccentricities of the Earth and Moon orbit, currently 0.017 and 0.0549 respectively. These are important and fun when figuring out when a TSE happens because the angular size of the Moon is so close to that of the Sun, the slight difference in distance between the Sun-Earth and Earth-Moon due to these eccentricities will change the angular size of the Sun or Moon. This is optimized for a TSE when the Earth is at aphelion, and Moon is at perigee, making the Sun its smallest and the Moon its biggest.

\subsection{An Orbital Argument to define TSE Frequency}\label{sec:orbits_modern}

\begin{figure}
\begin{center}
\includegraphics[width=0.6\linewidth]{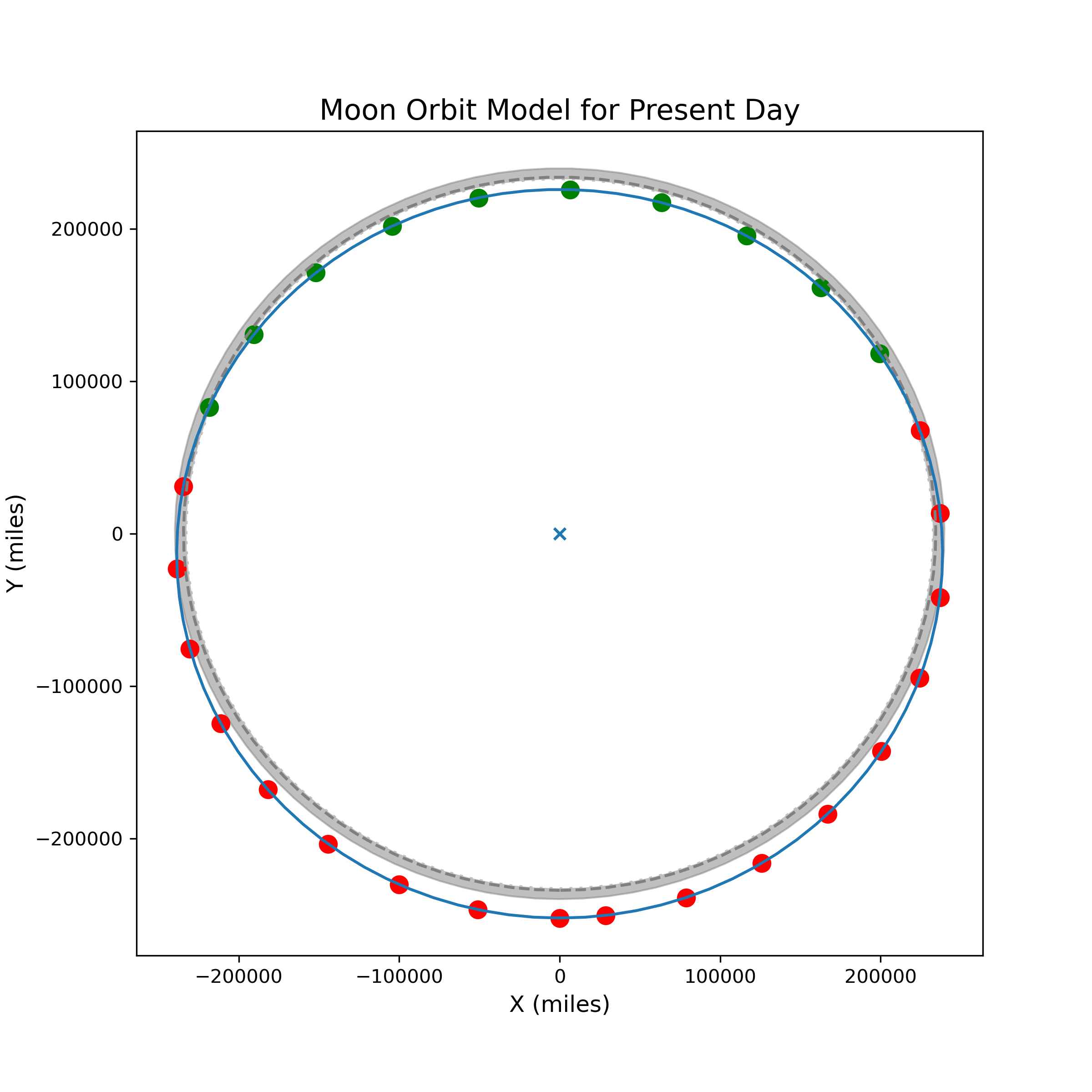}
\caption{A model of the Moon's orbit around the Earth for the present day. The blue line denotes the elliptical orbit. The gray dashed line is the maximum radial distance the Moon can be for their to be a TSE if the Earth is at its average distance from the Sun. The grey shell represents how much this radius changes as the Earth moves from aphelion to perihelion. Green points are sampled position on the orbit that are within the minimum radius, and red points are those outside of it. The blue x represent the Earths postion.}
\label{fig:orbit_modern}
\end{center}
\end{figure}

The common heuristic is that TSE currently happen roughly every 18 months. Not satisfied in doing things quickly, we made a simple model of the Moon's orbit (don't worry, we're gonna need this later). In Figure~\ref{fig:orbit_modern} we see the elliptical orbit of the Moon at its current semi major axis of 239000 miles. The grey dashed line is a maximum orbital distance from Earth for which the Moon's angular size allows for a Total Solar Eclipse when the Earth is its average distance from the Sun, approximately 234400 miles.

We then place representative ''Moons'' around the orbit, and note if they are within this minimum radius or outside of it. If we assume that each point in the Moon's orbit is equally likely to be aligned during syzysgy with the Earth and Sun, them the fraction of number of points within divided by the total number of points is a simple means of calculating how frequently a TSE occurs. In our simple model, 18 points are outside, and 10 are in, so if any apsis in the orbit is taken at random approximately 35$\%$ of them could be a TSE. However, there are 2 opportunities for syzygy in a year so at random there are 0.7 TSE per year, or one TSE every 1.4 years.

\subsection{Current World TSE Rate}

With our estimate roughly matching the heuristic for TSE frequency, we need to check if there is a location bias for TSE occurrence. \cite{meeus_location} used a statistical approach, tracking how often 408 standard points on the globe experienced an eclipse between 1799 and 2299 AD. In this 600 year span, they found that on average any given point on the earth experiences a TSE every 375 years. Furthermore, \cite{Wright_2024_website} took the NASA catalog of 5000 years of eclipses \cite{catalog} and tracked how often any of the 3743 Total or Hybrid eclipse paths landed on each of the $\sim$ few square mile pixels. They found a similar average of 366, mostly independent of longitude.

From these results we take two conclusions. In the 5000 years eclipse catalog, there were $\sim$ 3000 Total eclipses (Hybrid eclipses are edge cases we choose to ignore). This is a rate of 0.6 TSE per year, which roughly matches our value from Section~\ref{sec:orbits_modern}. Furthermore, the average of 366 years between TSE from \cite{meeus_location} and 375 from \cite{Wright_2024_website} inspire us to create a new term which we dub the Astronomical World Eclipse Surface cOverage MEtric (AWESOME) time. 

\subsubsection{AWESOME Time}\label{sec:awesome}

The AWESOME Time represents how long it takes for the entire surface of the Earth to experience a TSE on average. 
As a proof, we start by estimating the surface area of paths of totality from all the TSE in the \cite{catalog} catalog. Most TSE paths are 100 miles across, and stretch 10,000 miles from one end to the other across the curvature of the Earth's surface. Therefore, on average a TSE path covers $\sim$ 1 million square miles. The Earth's surface area is 196 square miles, so the 3000 TSEs from the 5000 years catalog covered the Earth's surface $\sim$ 15 times. If we divide the number of years to find how long on average it takes to cover the Earth's surface one time, we get an AWESOME time of 326 years.

We compare our work to others. Using the same catalog, \cite{Wright_2024_website} mapped all of the paths of totality and found the average number of times each pixel was on a path was 16 over the 5000 years, comparable to our 15. \cite{Wright_2024_website} calculated his own version of the AWESOME time at 375 years, and \cite{meeus_location} found a value of 366 years.

Satisfied with out definition, we must realize the AWESOME time is extremely powerful. Now we have a time period at which the entire globe will experience a TSE within, as long as we average it out over a much greater time period. This allows us to ignore the geographical ranges of animals as we can assume their entire range will experience on average one TSE at some point within an AWESOME time. Whether their population is located at a single point or dispersed evenly around the globe, by the end of an AWESOME time, on average their entire population will have experienced 1 TSE across the species.

\subsection{Lunar Orbital Retreat effects on AWESOME time}\label{sec:awesometime}

The lunar distance from the Earth will dramatically effect the AWESOME time over the course of Life's history. A closer Moon than today's will 1) have a TSE more often and 2) have a wider path of totality. 

To account for this, we use the orbital model from \ref{sec:orbits_modern} along with a linear estimation of the Moon's semi major axis evolution from \cite{moon_recession} of 3.8 cm/yr to calculate the Moon's semi major axis over time. This rate is consistent with more advanced models of lunar recession \cite[e.g.][]{ocean_main} over the billion years centered on today.

\begin{figure}
\begin{center}
\includegraphics[width=0.6\linewidth]{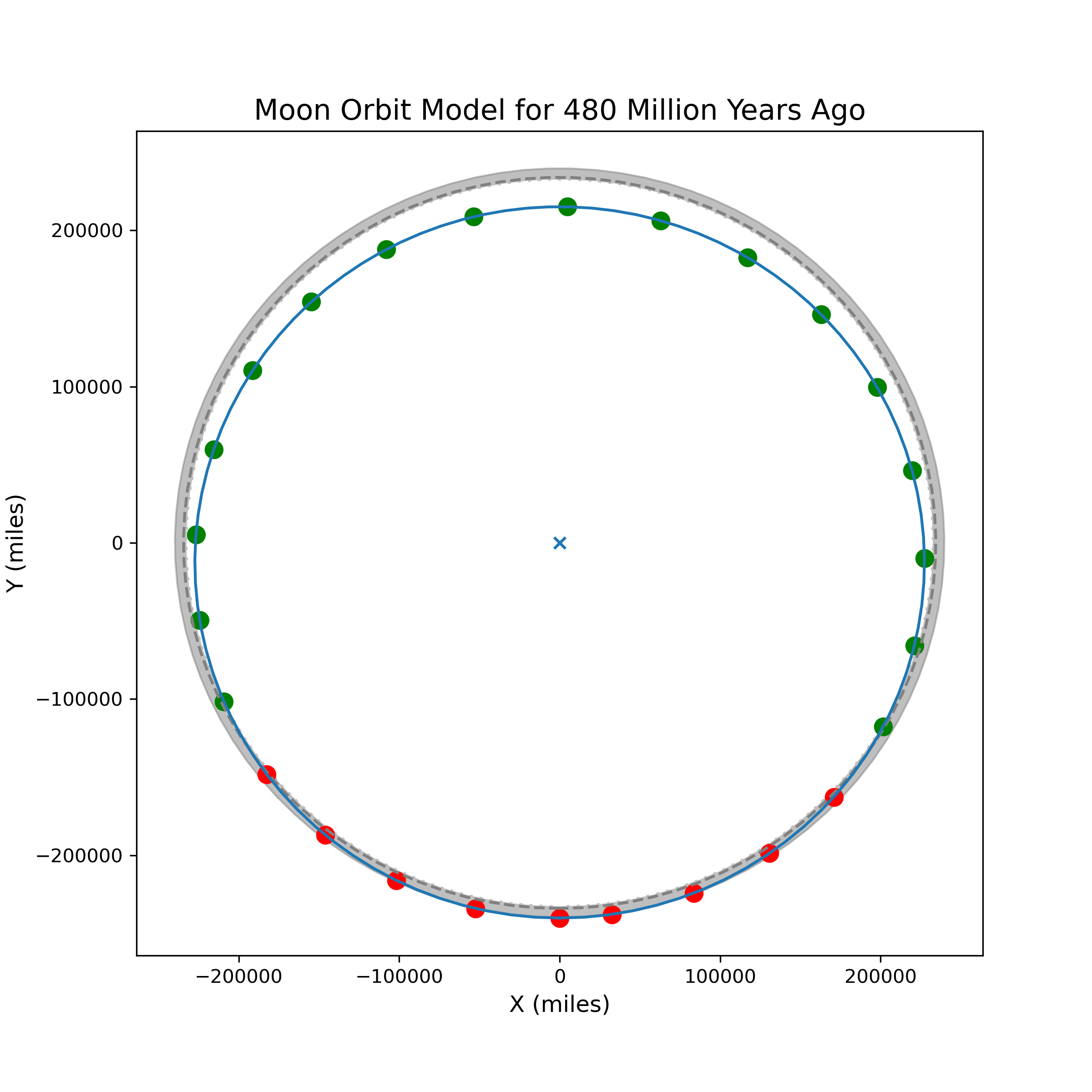}
\caption{Same as for Figure~\ref{fig:orbit_modern} but using a Moon semi-major axis for 480 million years ago. Notice the different number of red and green points.}
\label{fig:orbit_HS}
\end{center}
\end{figure}

Figure~\ref{fig:orbit_HS} shows the same set up as Figure~\ref{fig:orbit_modern} but places the Moon as the semi-major axis when Horseshoe crabs first appear in the fossil record 480 million years ago \cite{horseshoe_start}. Notice the frequency of total solar eclipses is far higher ($\sim$ 1.2 TSE per year). The much closer Moon means the Umbra and path of totality will be larger too. We estimate the Umbra radius on the Earth's surface using 

\[ r = R_m - \frac{d_m-Re}{d_e-d_m}*(R_s-R_m)\]

This is a geometric definition and doesn't account for the curvature of the Earth surface beneath the shadow, which can project the shadow over a greate distance. When we applied it to our modern model of the Moon orbit, we found we need to multiply the diameter of our Umbras by a fudge factor of 1.723 to match the $\sim$ 100 mile typical path of totality width. We apply the same fudge factor to totality paths at other epochs.

\section{Calculating TSE per Species}\label{sec:calculate}

We can now calculate the AWESOME time across the history of life on Earth. We divide up the last 500 million years in to 10 million year bins to calculate AWESOME times during. We also predict 500 million years into the future.

At each time step we make a Moon orbit model presented in Sections~\ref{sec:orbits_modern} and \ref{sec:awesometime}. From this we calculate the number of orbit positions within the minimum radius for a TSE, and calculate the time between eclipses (Sections~\ref{sec:orbits_modern}) and average Umbra shadow radius which we convert with the fudge factor in to a surface area covered by the path of totality (Section~\ref{sec:awesometime}). We then divide the Earth's surface area by the surface area of an average path of totality and multiply by the time between eclipses to arrive at an AWESOME time. 

\subsection{Results}

\begin{figure}
\begin{center}
\includegraphics[width=0.7\linewidth]{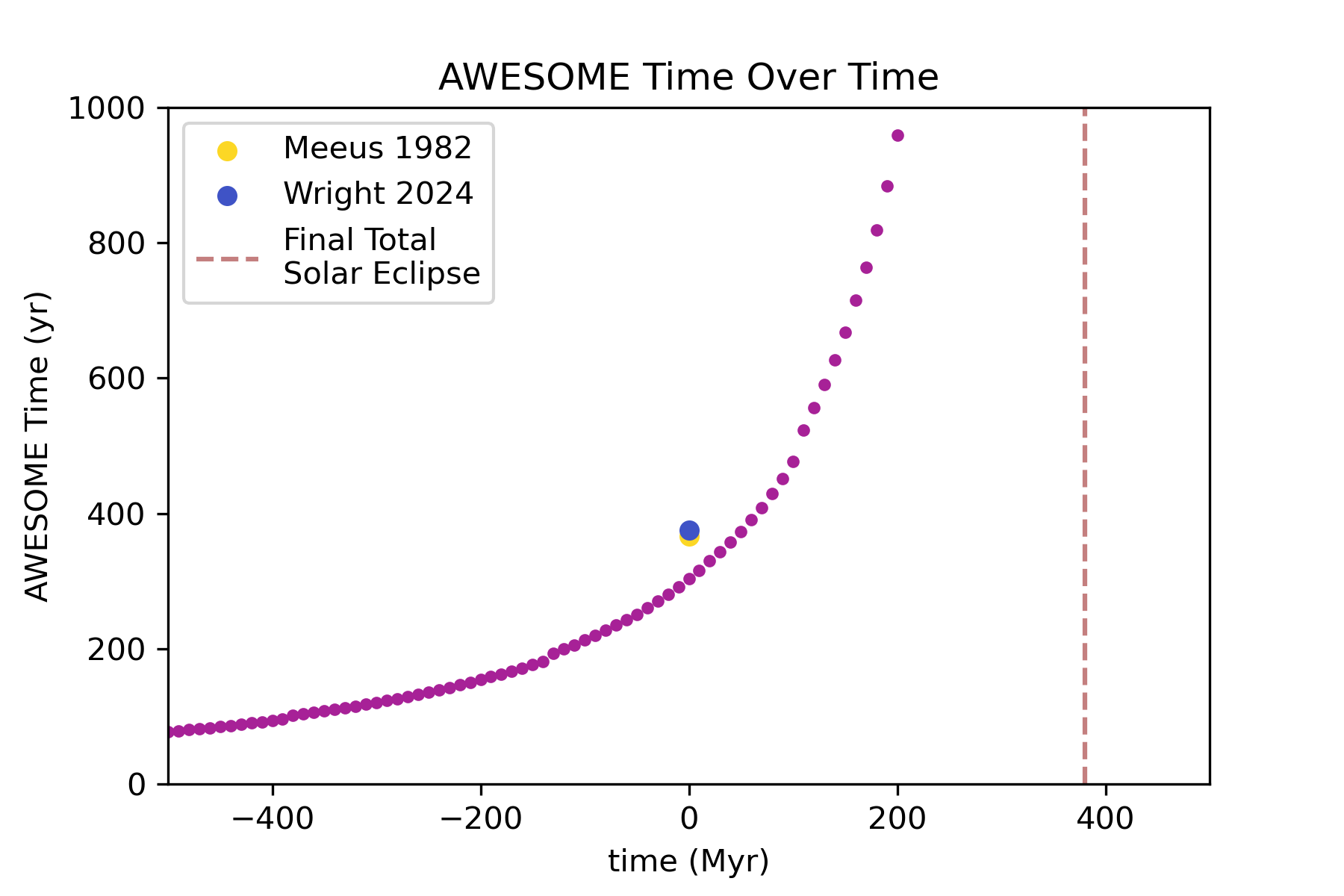}
\caption{A plot of AWESOME Time (the average time between Total Solar Eclipses for a point on the surface of the Earth) over 10 Myr periods in the 500 million years preceding and succeeding the present.}
\label{fig:awesome_time}
\end{center}
\end{figure}

We present the AWESOME time, the average time between Total Solar Eclipses for a spot on the Earth over 10 Million year increments of Earth's history in Figure~\ref{fig:awesome_time}. The AWESOME time was much shorter earlier in Earth's history. 500 million years ago the frequency of eclipses was an average of every 7 months with 200 mile wide paths of totality. On average it took 77 years for every place on the Earth to experience a True Solar Eclipse.

The AWESOME time exponentially grows over the course of the billion years displayed. In 370 million years, it will take 56000 years on average between successive coverings of the Earth by Total Solar Eclipses. The time between eclipses will be 7.5 years on average, but the width of the path of totality will only be $\sim$ 3 miles wide. The final Total Solar Eclipse will occur approximately 380 million years from now.

\begin{figure}
\begin{center}
\includegraphics[width=0.7\linewidth]{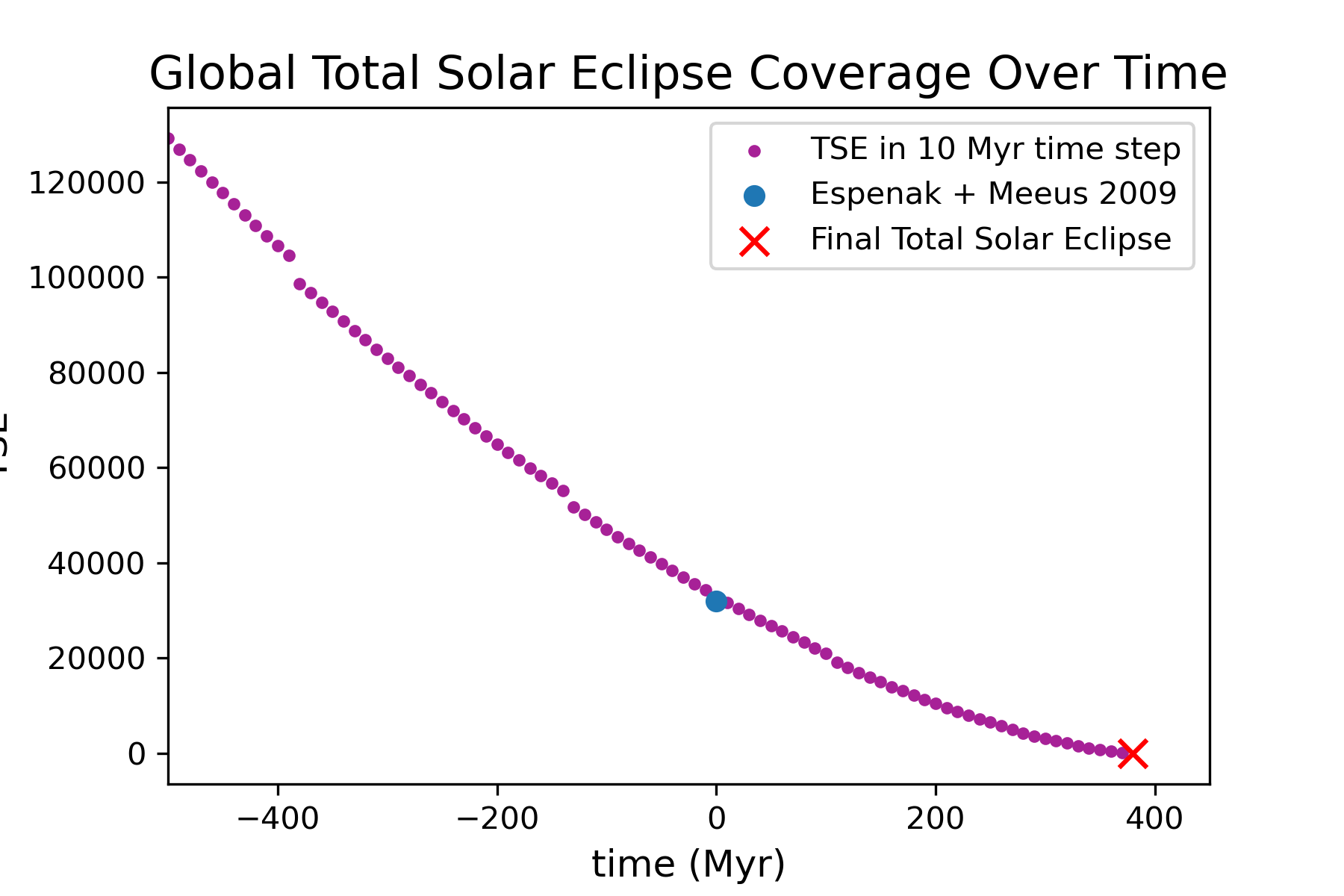}
\caption{A plot of the number of Global Total Solar Eclipses Coverages for the Earth during a 10 Myr period in the 500 million years preceding and succeeding the present. This is how many times the surface of the Earth will be covered by TSE, and also the average of how many TSE a single point on the Earth will experience in that 10 Myr period.}
\label{fig:tse_time}
\end{center}
\end{figure}

In Figure~\ref{fig:tse_time} we present the number of TSEs a point on the Earth will experience on average during 10 million year bins in Earth's history, or how many times the Earth will be wrapped in TSE during that time. We do this by dividing the 10 million year bins by their AWESOME Time. We see the number of eclipses drop exponentially, until at 380 million years in the future there is a final Total Solar Eclipse, and then no more.

\subsubsection{Horseshoe Crabs and Humans}

\begin{figure}
\begin{center}
\includegraphics[width=0.7\linewidth]{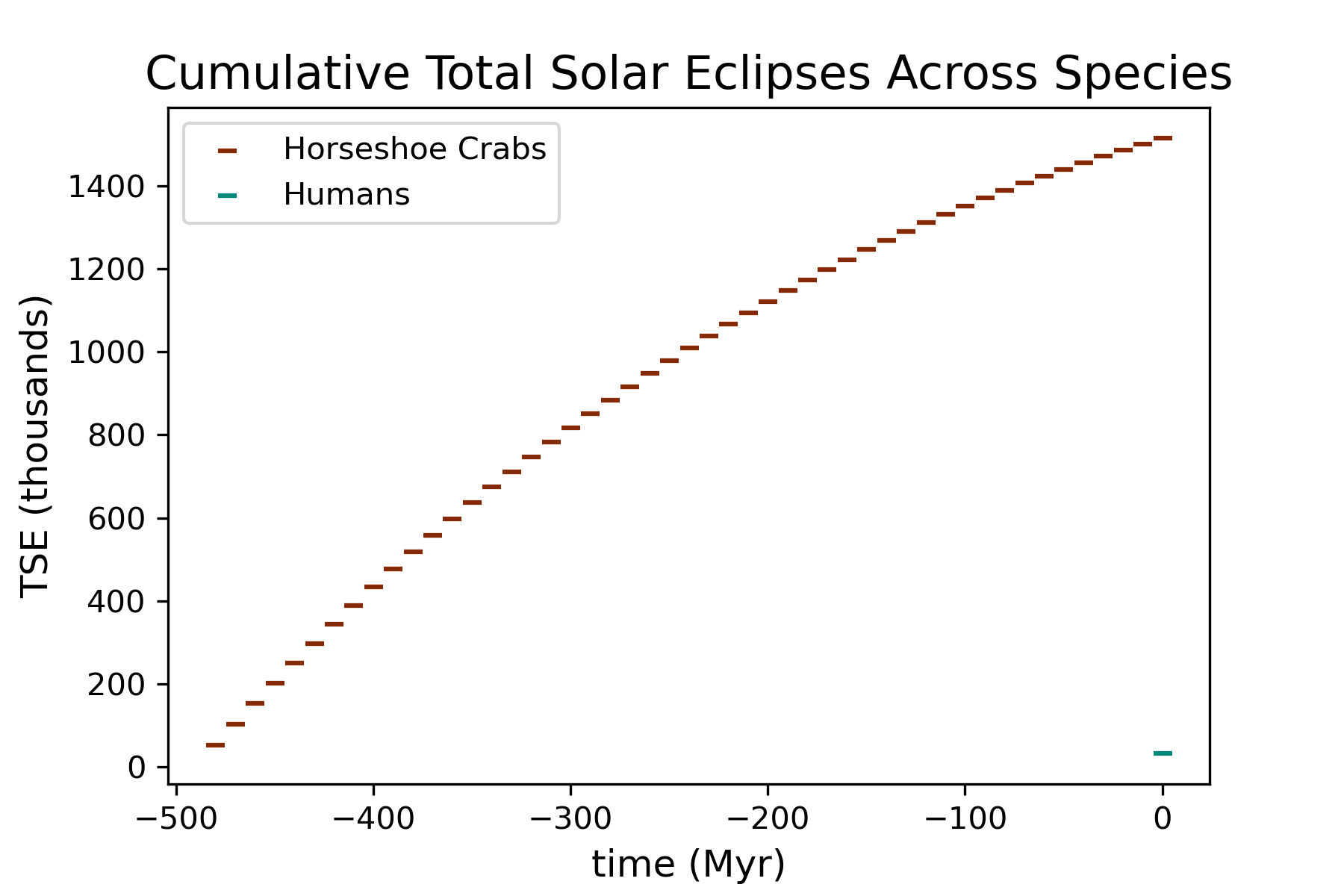}
\caption{A plot of the cumulative number of Total Solar Eclipses the family \textit{Limulidae}, Horseshoe Crabs has experienced. Starting from 480 million years ago, the number of TSE in Figure~\ref{fig:tse_time} in each bin is added to the total. A factor of 5/12 is also applied, representing that only 5 of the Months of the year are Horseshoe Crabs attempting to mate on the beaches and able to observe a TSE. The total number of TSE for Humans is also presented.}
\label{fig:cumhorse}
\end{center}
\end{figure}

In Figure~\ref{fig:cumhorse} we present the cumulative number of TSEs observed by Horseshoe Crabs and Humans. It's important to note that we attempted to factor in Horseshoe Crab behavior into this result. Horseshoe Crabs leave the ocean floor and spend time on the beach to mate over the mating season, March through July \citep{horseshoe_mating_season}. We therefore applied a factor of 5/12 to the totals in Figure~\ref{fig:tse_time} to account for the months out at sea. Horseshoe crabs time their beach rendezvous with the spring tide caused by the New and Full Moon, so it is not unrealistic that they are likely to be on the beach during TSEs that fall during mating months.

To date, the family \textit{Limulidae} has witnessed 1.5 million eclipses across the species. This number is not the total that have happened, but given a dispersed population, their entire range has been covered that many time. Therefore, we can multiply this number by an assumed steady population and get a number for the total number of individual eclipse experiences. Good Horseshoe population data starts in what must be assumed is the middle of a massive drop off (see Section~\ref{sec:ani_pop}) but in 1871 4 million Horseshoe Crabs were ''harvested'' in Delaware and New Jersey. We can assume the population was much higher before records of fishing were kept, and while this is a major breeding ground for Atlantic Horseshoe Crabs, they do spread from Maine to Mexico. Therefore we estimate a conservative standing population of 40 million Horseshoe crabs for that species, and naively assume the same for the other 3 extant species.

If a standing population of 120 million Horseshoe Crabs witnessed each of the 1.5 million eclipses, that would make a total of 138 trillion Total Solar Eclipse experiences for Horseshoe Crabs.

In comparison, Humans have witnessed just 32000 TSE across the species, and that includes extending the definition of Humans back to the last common ancestor of Humans and Chimpanzees 10 million years ago. Human population numbers have increased exponentially in the last 10,000 years, but prior to that were on the order of tens to hundreds of thousands \citep{human_pop}. 

We assume a standing population of 1 million to keep things simple, meaning across the Human species there have been 32 billion TSE experiences.

\section{Discussion}\label{sec:disc}

Given that Horseshoe Crabs have experience over 3 orders of magnitude more TSEs than Humans, can we catch up? The 2017 Total Solar Eclipse had an estimated 215 million participants. If humanity can mobilize that number of people to each upcoming eclipse (including those over the ocean) we can overtake Horseshoe Crabs in $\sim$ 400000 years.

This does not seem feasible however, and if we return to the approach above and set a new standing population of 10 billion individuals and predict the number of eclipses across species, we would overtake the current Horseshoe Crab TSE experiences in under 10 million years.

But at what cost does that Human population entail? Although a potential synthetic replacement is being developed, the population of Horseshoe Crabs continues to suffer due to harvesting for their blood as an important tool in pharmaceuticals \citep{new_horse}. The Delaware population is struggling to maintain 700000 members, down from estimated tens of millions. Our global civilization exists due to the exploitation and extermination of resources and environments that have sustained other species for millions of years. A total solar eclipse is worth sharing.

\section{Conclusions}\label{sec:con}

The AWESOME time is the average time between TSEs for a given point on the Earth, as well as describing the average time it takes the whole Earth to experience one TSE.

Total Solar Eclipses have evolved over the Earth's history, and the final Total Solar Eclipse will be in approximately 380 Million Years. If we want more eclipses after that, we will need to move the Earth, Moon or Sun.

We have presented a framework to calculate the number of Total Solar Eclipse experiences for any given species in Earth's history.

We found that Horseshoe Crabs across the species have had 138 Trillion Total Solar Eclipse experiences, compared to Humans with only 38 Billion. However, if population levels maintain, Humans will overtake them in less than 10 million years.

\begin{acknowledgments}
The author would like to thank J.W., E.G., W.D., D.V., and A.M. for humoring him 7 years ago and agreeing to go witness his first total eclipse. He would also like to proactively thank all those he will be dragging to future ones.

\end{acknowledgments}

\bibliography{bilbo}{}
\bibliographystyle{aasjournal}

\end{document}